## 1. Title:

The Interaction between PEDOT/PSS Gate and sub-μ low-k Insulator for All Printed OFETs

## 2. Authors and their Affiliations :

Z. Shalabutov*, Institut für Print- und Medientechnik, TU Chemnitz, D-09126 Chemnitz, Germany
M. Friedrich, Institut für Physik, TU Chemnitz, D-09107 Chemnitz, Germany
I. Thurzo, Institut für Physik, TU Chemnitz, D-09107 Chemnitz, Germany
A. Hübler , Institut für Print- und Medientechnik, TU Chemnitz, D-09126 Chemnitz, Germany
D. R. T. Zahn, Institut für Physik, TU Chemnitz, D-09107 Chemnitz, Germany
U. Hahn , Institut für Print- und Medientechnik, TU Chemnitz, D-09126 Chemnitz, Germany
* Corresponding author. Tel. +49-371-531-2376; *E-mail address*: szl@hrz.tu-chemnitz.de

### Abstract

In all printed OFETs with Poly(3,4-ethylenedioxythiophene)/ poly( styrenesulfonate) (PEDOT/ PSS) gate, offset printing and gravure of electrically dense sub-μ insulators from polyvinylphenol (PVPh), polyvinyl alcohol (PVOH) and poly(methyl methacrylate) (PMMA) as well as other organic and inorganic materials turned out to be problematic due to the most reactive part of the gate material- the negative sulfonate ions from PSS. The present paper investigates the nature of the interaction between PSS and PVOH sub-μ insulator explored by infrared spectroscopy and electrical methods. Some evidence is obtained that most probably $OH^-$ and not sulfonate ions are responsible for creating channels, penetrated by PEDOT/PSS nano dispersion applied as gate.

## 3. Text of paper

### Introduction

During the last few years, the performance of organic field effect transistors (OFETs) has improved to a point of being viable in industrial applications [1, 2]. Polymers represent a cost efficient alternative to expensive silicon based products. Commonly used techniques like sublimation of single crystals or vacuum deposition of small molecules however are not cost effective for inexpensive mass production. A more suitable approach is solution processing of polymers or precursor materials, which can be done by spin-coating, doctor-blading, or printing. Printing is most preferable because patterned layers are directly deposited in a fast and efficient way.

In order to observe the field effect in OFET, there are three possibilities for the insulator:

**a/** low-k [3,4] – insulating layer with sub-μ thickness with low dielectric constant

**b/** high-k – insulating layer with thickness more than 5-6 μm with high ( around 20) or even very high dielectric constant (higher than 1000)

**c/** combination of low-k and high-k materials- in order to eliminate or reduce ion trapping effects , typical for the high-k material.

While constructing OFETs, some authors [5, 6, 14 ] use PEDOT/PSS as the gate material. The doped PEDOT with PSS is a very well studied polymer and because of its excellent submicron film forming properties, it is adequate to be used as gate and S/D structure in printed polymer electronics.

As gate- applications techniques screen printing [7-9], micro contact printing [10,11], inkjet printing [12-14], pad printing [15], UV lithography [16] were used.

An often occurring problem when constructing all printed OFETs with low– k submicron dielectric and PEDOT / PSS gate is that after the gate deposition short circuits are detected. Sometimes this is attributed to defects like pinholes, craters, etc. and this problem is eliminated by applying several layers till no short circuit is measured [12] or by modifying the insulator with adequate additives. In other cases, one can think of PSS strong reactive negative ions attacking and destroying the insulator.

The investigation of causes leading to short circuits, still remains a challenge. On the other hand, at present there is a large amount of literature concerning conducting, semi- conducting parts and surface reactions in OFETs, but to our knowledge, there is no condensed information dedicated to studying the nature of the interaction (not only on surface, but also in volume) between PEDOT/PSS gates and the insulator in OFETs.

Realising such a study can open possible future ways to develop electrically dense submicron low-k gate insulators for all printed electronics.

## Experimental

In order to investigate the nature of the interaction between the PEDOT/PSS gate and the PVOH submicron dielectric layers, on p-Si- wafers using spincoating techniques, we deposited first the dielectric and second PSS 1% above was dispensed.

Infrared transmission spectra were recorded using a Fourier transform infrared spectrometer Bruker IFS 66. PSS and PVOH were deposited onto double sided polished 1 mm thick Si wafers. The Si wafers are transparent in the mid infrared spectral range. All measurements were done with a spectral resolution better than 2 $cm^{-1}$. For a better comparison all measured spectra are presented in absorbance units. In the following figures the logarithm of the transmission ratio of the bare and film covered substrate after baseline correction is presented.

Electrical characterization in order was based on measurements of cyclic voltammetry (ZV) and admittance: capacitance-voltage (CV), conductance-voltage (GV) [17]. An attached Pt-electrode ($\varphi \approx 0.1$ mm) was used as the top contact while the p-Si substrate served as the bottom electrode to obtain hints on ion motions and electrode reactions in the PSS/PVOH/p-Si system.

## Results and discussion

Fig. 1 shows in the upper part the infrared spectrum of a thin PVOH layer and in the lower part the spectrum of the same film after deposition of PSS.

In the upper spectrum characteristic bands of PVOH are displayed. These are at 3340 $cm^{-1}$ and at 2900 $cm^{-1}$ the stretching vibration bands $\nu$(OH) and $\nu$(CH), respectively. In the low wavenumber range between 1300 $cm^{-1}$ and 1500 $cm^{-1}$ $\delta(CH_2)$, $\gamma_\omega(CH_2)$ and $\delta$(CH+OH), at 1085 $cm^{-1}$ $\nu$(OH) and at 850 $cm^{-1}$ $\gamma_r(CH_2)$ appear [18].

The PVOH/PSS spectrum after the PSS deposition is shown in the lower part of fig. 1. The low wavenumber range is dominated by the PSS spectrum. Two additional features due to $CO_2$ contribution in air appear between 2200 and 2400 $cm^{-1}$ and at 670 $cm^{-1}$. In order to distinguish between the spectra of the two components the upper spectrum of the PVOH layer was subtracted from the lower spectrum. Additionally the spectrum from a single PSS sample was taken. The results are presented in fig. 2. As can be seen in the figure the upper difference spectrum is in good agreement with the lower spectrum of pure PSS. Below 3000 $cm^{-1}$ peak positions coincide. Furthermore no additional bands are observed. Deviations in the half widths are caused by thickness inhomogeneity of the coverage. From the infrared results it can be derived that no strong chemical reactions take place in the PVOH/PSS system. Deviations are observed in the spectral range above 3000 $cm^{-1}$. The different OH band shape at 3400 $cm^{-1}$ indicates changes in the concentration and bond strength of OH groups.

In Fig. 3 the time behaviour of spectral changes taking place in the PVOH/PSS sample is displayed. The decrease of both peaks at 1640 $cm^{-1}$ and at 3400 $cm^{-1}$ which are characteristic for water indicates that the system looses water. With increasing delay time the OH peaks at 3400 $cm^{-1}$ shift to higher wavenumbers indicating the stronger bonding of remaining OH groups. Beside strong changes in the OH peaks due to water loss also small changes are observed in the spectral range below 1500 $cm^{-1}$. The most pronounced changes in the lower spectral range were detected at ~1050 and ~1260 $cm^{-1}$. For a detailed evaluation further systematic investigations are required.

Summarising the results from infrared spectroscopy (IRS) fig. 1 and 2 confirm that no chemical reactions between PVOH and PSS occur. The comparatively slight changes in OH band position and intensity displayed in fig. 3 are mainly caused by changes in the water content of the PVOH/PSS sample.

In order to interpret the results from IRS a set of electrical experiments were carried out with emphasis on transient charging.

To characterize charge transport, first of all ZV-Measurements were conducted using Pt as the top and p-Si as the bottom electrode. A typical dependence of the current I on the potential $U_{Pt}$ of the Pt-electrode is shown in figs. 4 and 5.

The remarkable hysteresis appears for the 5V→-5V half-cycle ($\delta U$ = -0.1 V), the potential $U_{Pt}$ of the current reversal is shifted by ≈ 1.3 V (offset) away from 0 V. This phenomenon is a result of the competition between the steady-state (*dc*) current $I_0$ and the transient current $I(\delta t)$. The symbol $\delta t$ stands for the delay of a current reading after the potential has been changed by the step $\delta U$. Actually there is a transient current for $\delta U < 0$ only, as documented by Fig. 6. The steady-state current is always zero if $U_{Pt}$ = 0 V. Is there no transient current $I(\delta t)$, which is the case of $\delta U > 0$, then even the total current is zero at zero bias. According to Fig. 6 for $\delta U < 0$ and $U_{Pt} > 0$ the *negative* transient current is compensated by the *positive dc* current $I_0$ ($U_{Pt} \approx 2$ V). Just the opposite is true for both $U_{Pt}$ and $\delta U$ becoming negative, there is an excess total current due to the summation of the two contributions. The presence of the negative transient current at $U_{Pt} = 0$ V und $\delta U = -0.1$ V is also documented by the difference I(0.5 s) -I(5 s), the corresponding decay time constant of I(δt) being much higher than 0.1 s. There is no significant difference in the respective behaviour of PVOH and PSS/ PVOH configurations. The asymptotes (dashed lines) in Fig. 6 demarcate an irreversible electro-chemical reaction resolved as the peak at $U_{Pt} \approx -0.9$ V.

The hysteresis and the positive offset of the potential for I = 0 (Fig. 3, $\delta U = -0.1$ V) are treated in literature alternatively as due to either a slow capture at deep traps [19] or originating from recharging an ion double layer [20]. Yet in our case they can be indications for the transport of negative charges that can be either sulphonic anions from PSS or $OH^-$ formed at the surface of PVOH sub-μ film due to the interaction of PVOH with air humidity.

Comparing the voltammetry curves of Pt/PVOH /p-Si (Fig. 5) and Pt/PSS/ PVOH/ p-Si (Fig. 4), they are almost identical, indicating that the detected movement is primarily from $OH^-$. This fact complies with the results from IRS. Of course, it is possible that there is a low contribution of the sulfonate ions of PSS to the process as well.

During a sequence of whole cyclic voltammetry runs, progressively lowered values for the total electrical current were measured every time. This can be explained by the irreversible redox reaction of the mobile ionic species.

To get a deeper insight into the observed anomaly, frequency-domain admittance measurements (CV, GV) were performed. In the frequency range from 10 kHz to1 MHz the equivalent parallel capacitance C is independent of frequency or bias (potential $U_{Pt}$). However, the equivalent parallel conductance G, which corresponds to dielectric loss, shows a clear dependence on frequency – cf. Fig. 7. Below 10 kHz it was impossible to detect the dielectric losses. Above this frequency the conductance follows the power law

$$G \propto \left( \frac{\omega}{\omega_0} \right)^n, \quad n \approx 1.$$

Replacing G by the *ac* conductivity $\sigma_{ac} \propto \omega\varepsilon''$, we arrive at the conclusion of dealing with a *constant* dielectric loss ε''. Such a behaviour of ε'' is described in topical literature as „*nearly constant loss* (NCL)“ [21-27]. As to the mechanism of the NCL, (local) ion movements are mentioned mostly for glasses and oxides. The assumption of having to do with ion motions in the case of PVOH is supported also by the observation of a gradual decrease of the total current on repeating ZV cycles (Fig. 4). This is true for structures without a deposited PSS as well.

As a deduction from IRS and electrical experiments, a hypothetical model has been created for a PVOH sub-μ insulating layer, to be applied to OFETs. Due to interactions of PVOH with the air humidity, $OH^-$ negative ions are adsorbed on the surface of the dielectric layer. They diffuse and after applying a negative gate bias migrate (mass transport) in the direction towards the S/D structure. A part of them remain trapped in the insulating film. Those ions which reach the S/D structure are neutralized and thus the net number of the mobile electrical charges decreases (by every cycle, lower total current is measured).

The charges, crossing the dielectric layer and neutralizing on the S/D structure, leave empty sites behind. After the deposition of PEDOT/PSS nano-dispersion, the conductive material enters the empty sites and, after drying, conductive channels connecting the large-area gate with S/D structures are formed. After applying a gate voltage, short circuits are observed.

## Conclusions

The nature of interaction between PEDOT/PSS and PVOH sub-μ layers was investigated using infrared spectroscopy and electrical measurements. No strong chemical changes in the double layer devices are observed by infrared spectroscopy. It was shown that the charge transport in PVOH sub-μ layer after deposition of PSS is dominated by $OH^-$ ions moving towards the p-Si substrate.

Slight changes in the infrared difference spectra most likely related to the water influence support the proposed transport mechanism.

The electrical measurements showed that the system PEDOT/PSS - PVOH sub-μ layer is homogeneous (free of pinholes) for attached electrodes with a diameter of 0.1 mm or smaller.

## Figure captions

Fig.1: Absorbance spectra of a PVOH layer (upper part) and of the same layer after partcial covering with an additional PSS (lower part).

Fig. 2: The residual absorbance spectrum obtained after substraction of the spectrum of the PVOH layer from the PVOH- PSS double system spectrum (upper part) is presented in comparison to the spectrum of a single PSS layer (lower part).

Fig. 3: Difference of the absorbance spectra of a PVOH-PEDOT/PSS double layersample to the initial spectrum measured after different delay times (red: 25 min, black: 80 min) to the initial spectrum. The difference of spectra after 25 and 80 min delay time is displayed below in green colour. The thin perpendicular grey dotted line marks the maximum position of the OH stretching band of the green curve. It is obviously that the maximum is shifted to higher wavenumbers.

Fig. 4: Two consecutive I-$U_{Pt}$ cycles point to a well resolved hysteresis of characteristics. Each cycle started and ended at $U_{Pt}$ = -5 V, the potential was scanned in $\delta U$ = 0.1 V-steps. There was always a delay $\delta t$ = 0.5 s before the next reading of the current.

Fig. 5: Two successive cyclic voltammetry runs starting at –5 V in each case for the Pt/PVOH/p-Si structure.

Fig. 6: The differential current $\Delta I$ = I(0.5)-I(5) corresponds to two ZV -measurements of a Pt/PVOH/p-Si structure with $\delta t$ = 0.5 s and 5 s, respectively.

Fig. 7: The conductance G, proportional to dielectric losses $\varepsilon''$, is a linear function of frequency f.

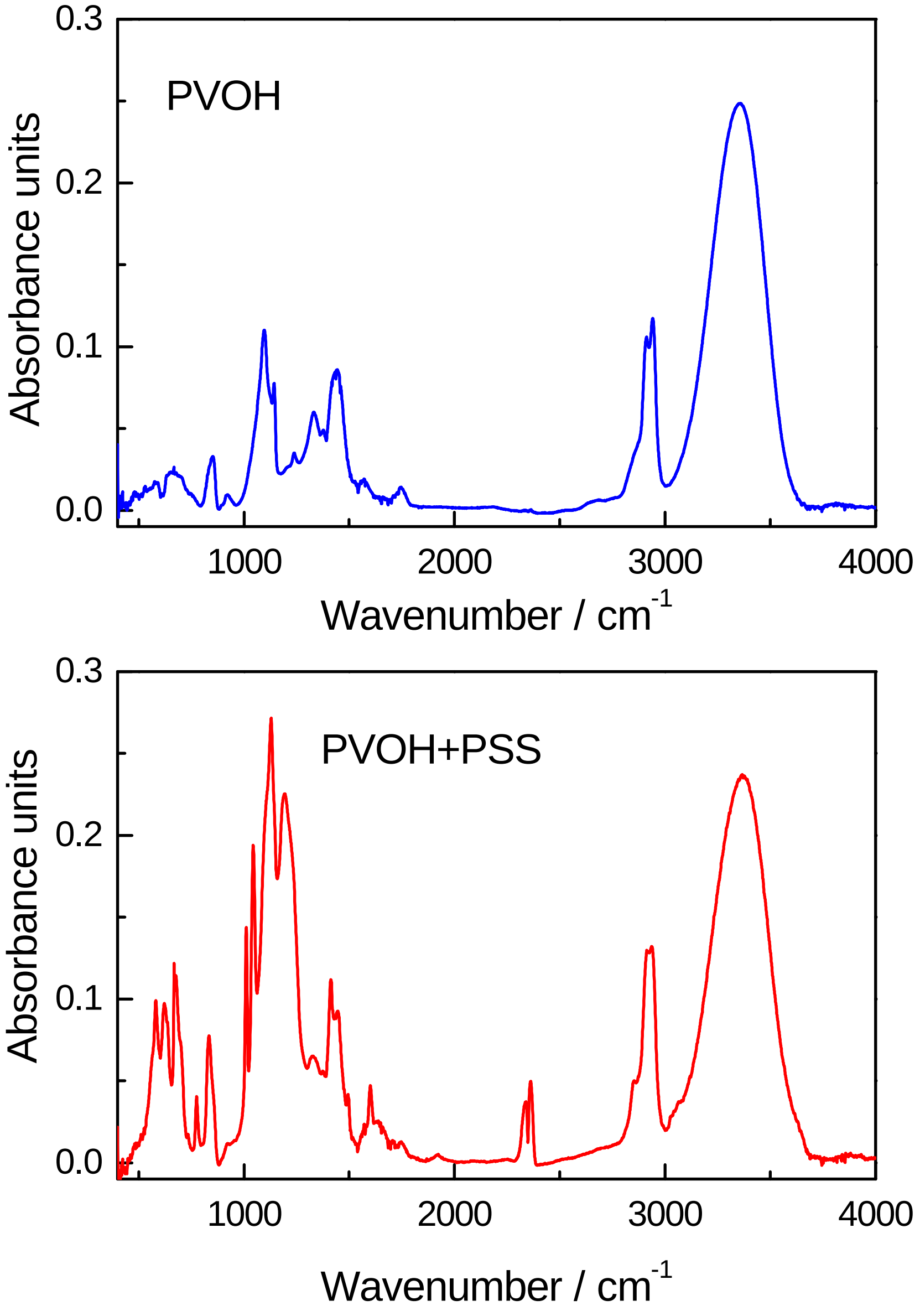

Fig. 1

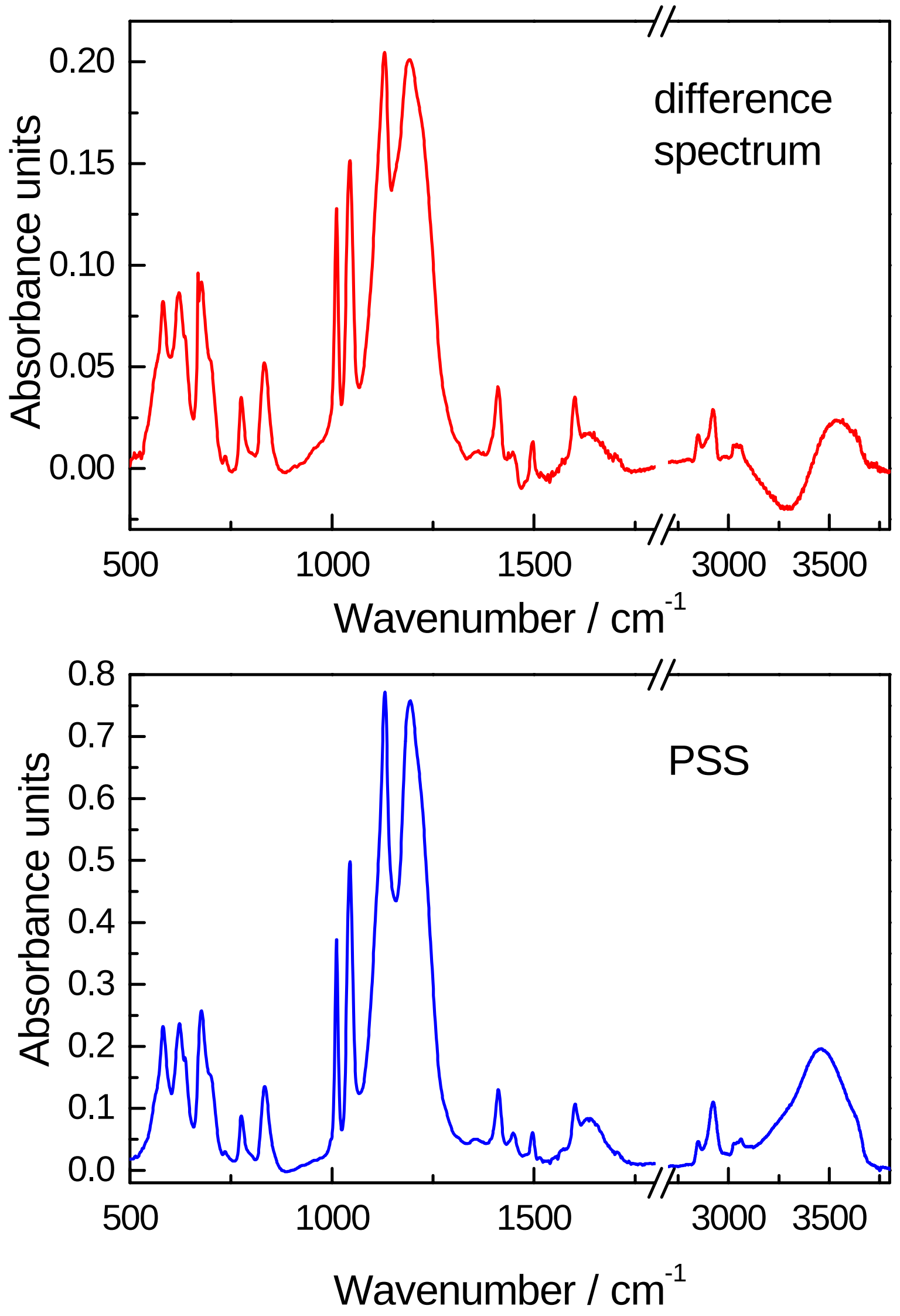

Fig. 2

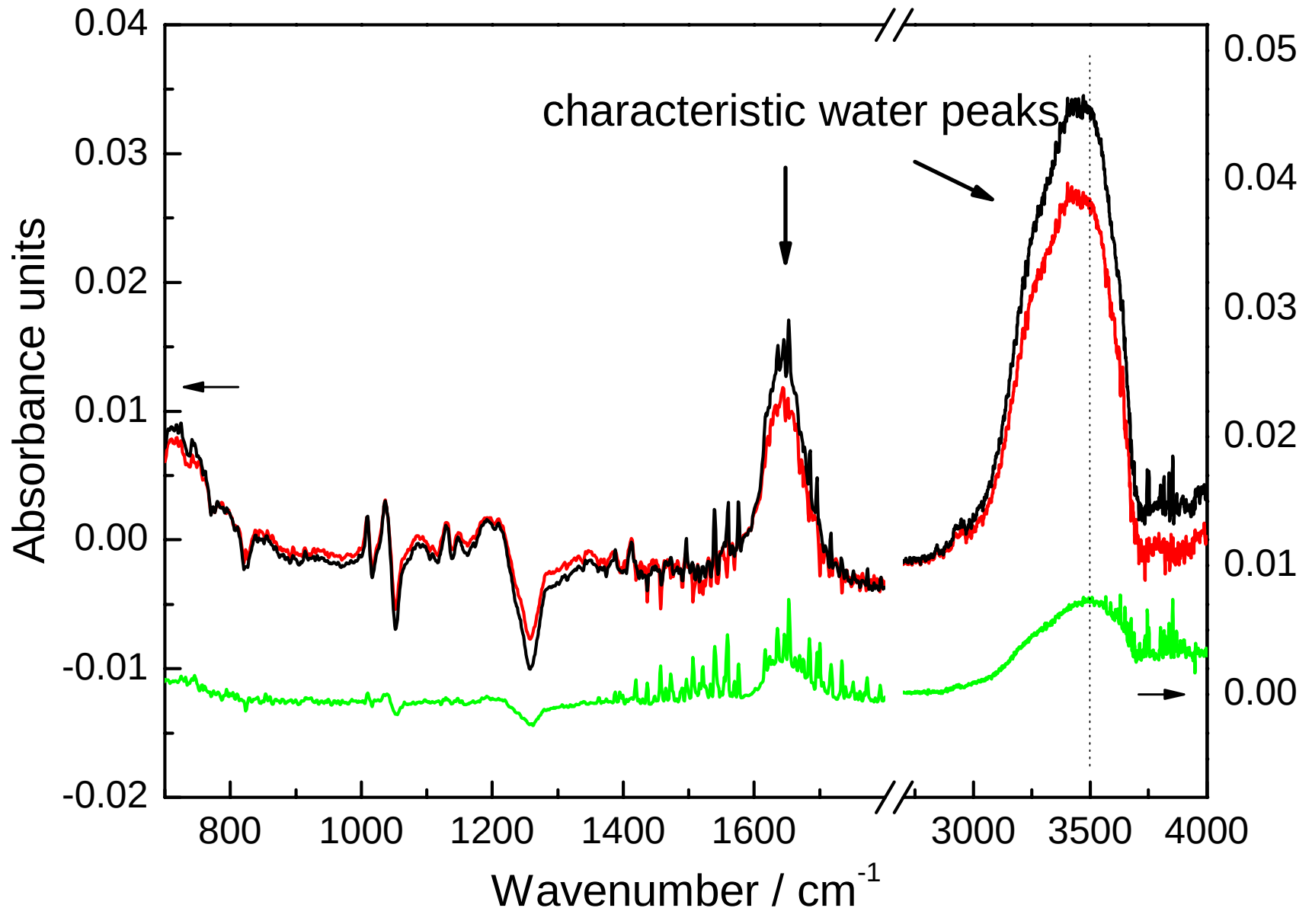

Fig. 3

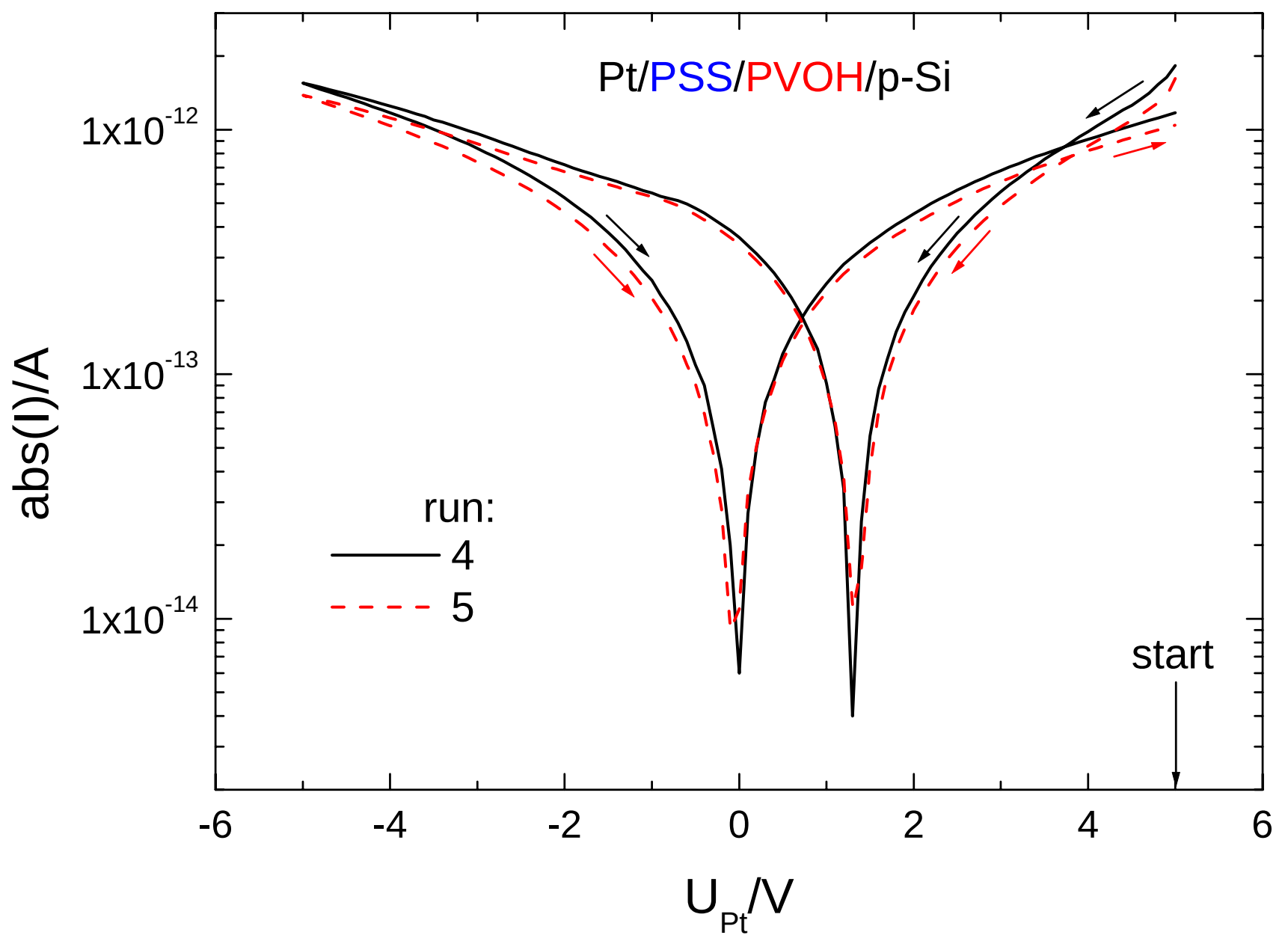

Fig. 4

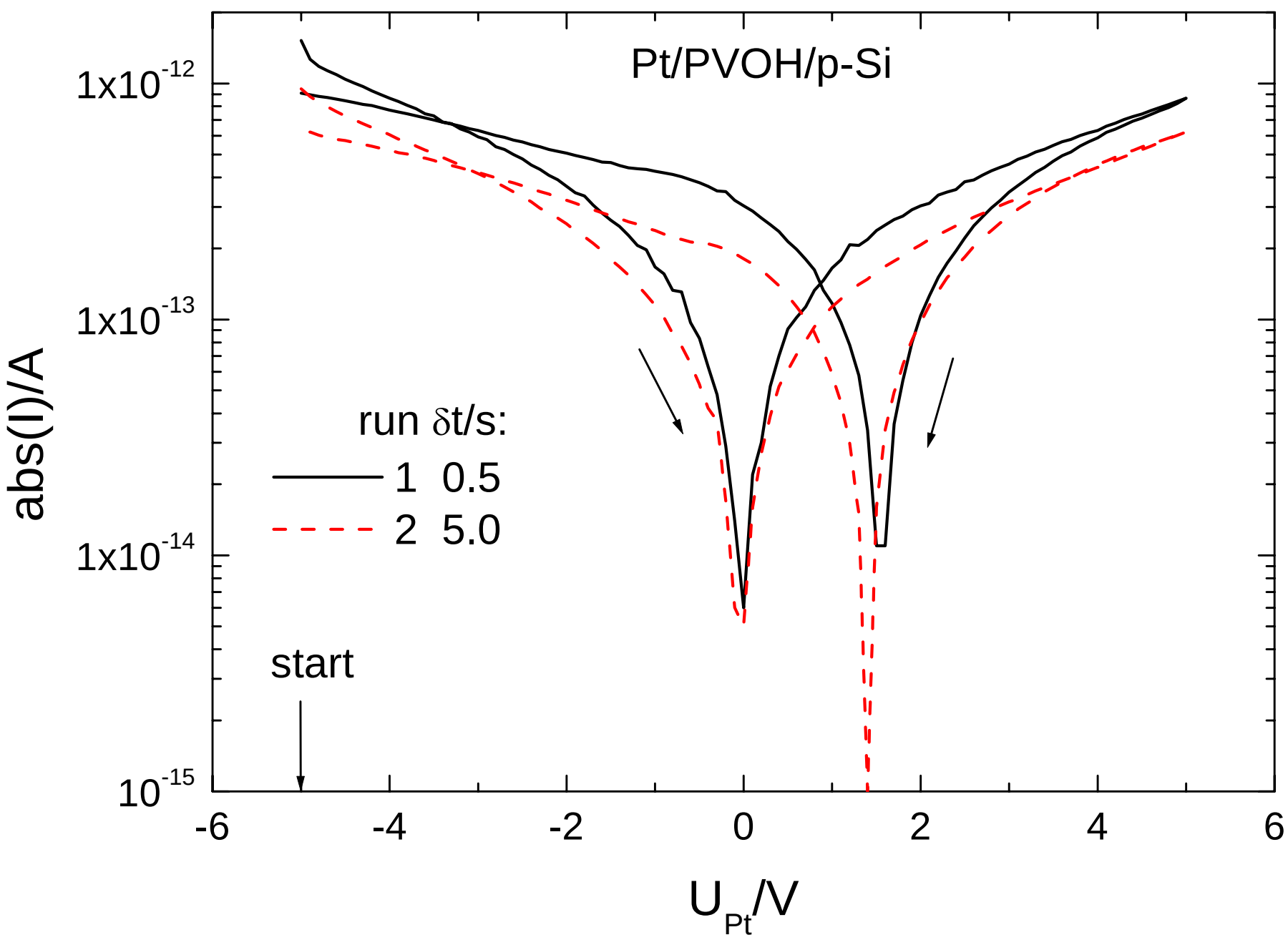

Fig. 5

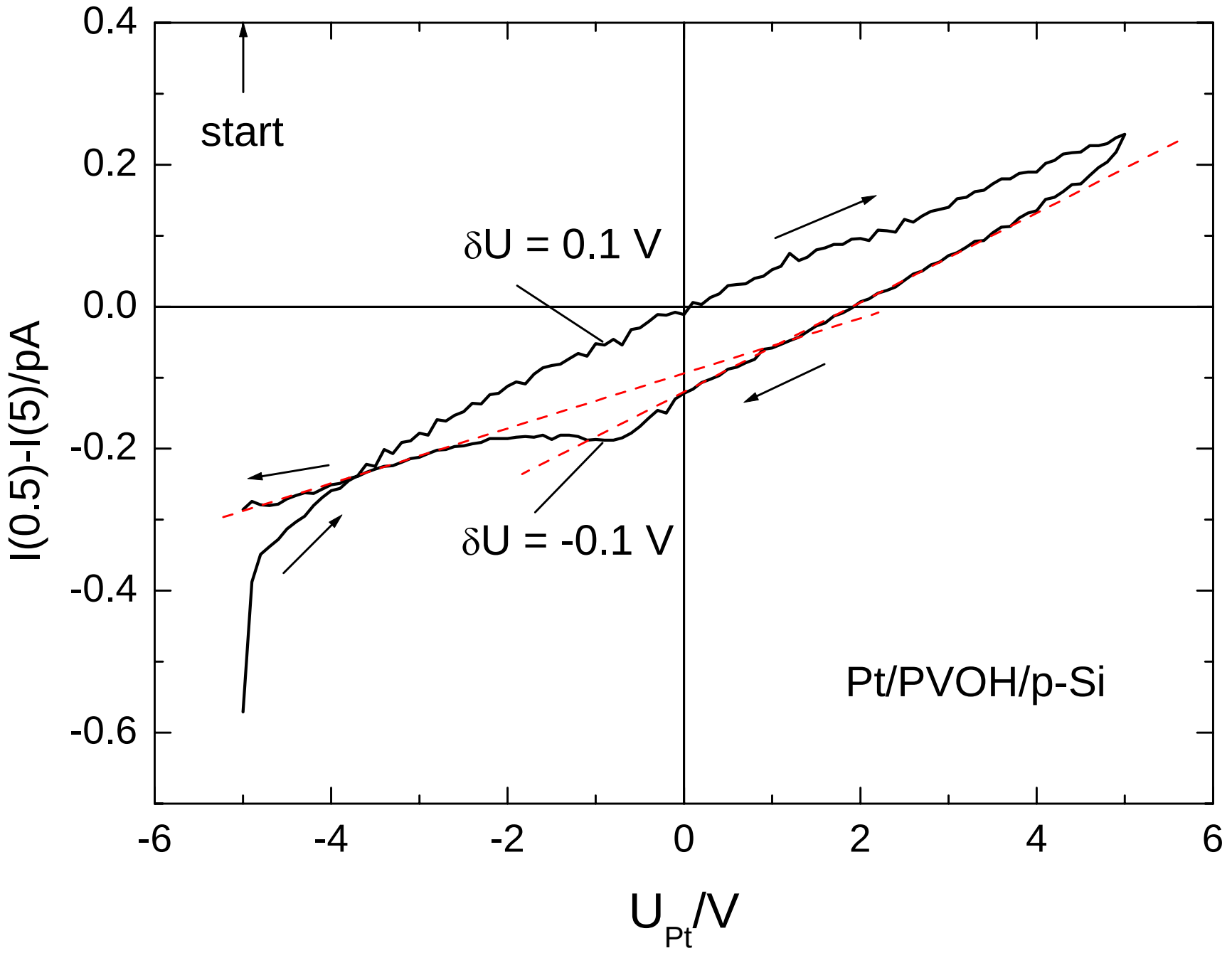

Fig. 6

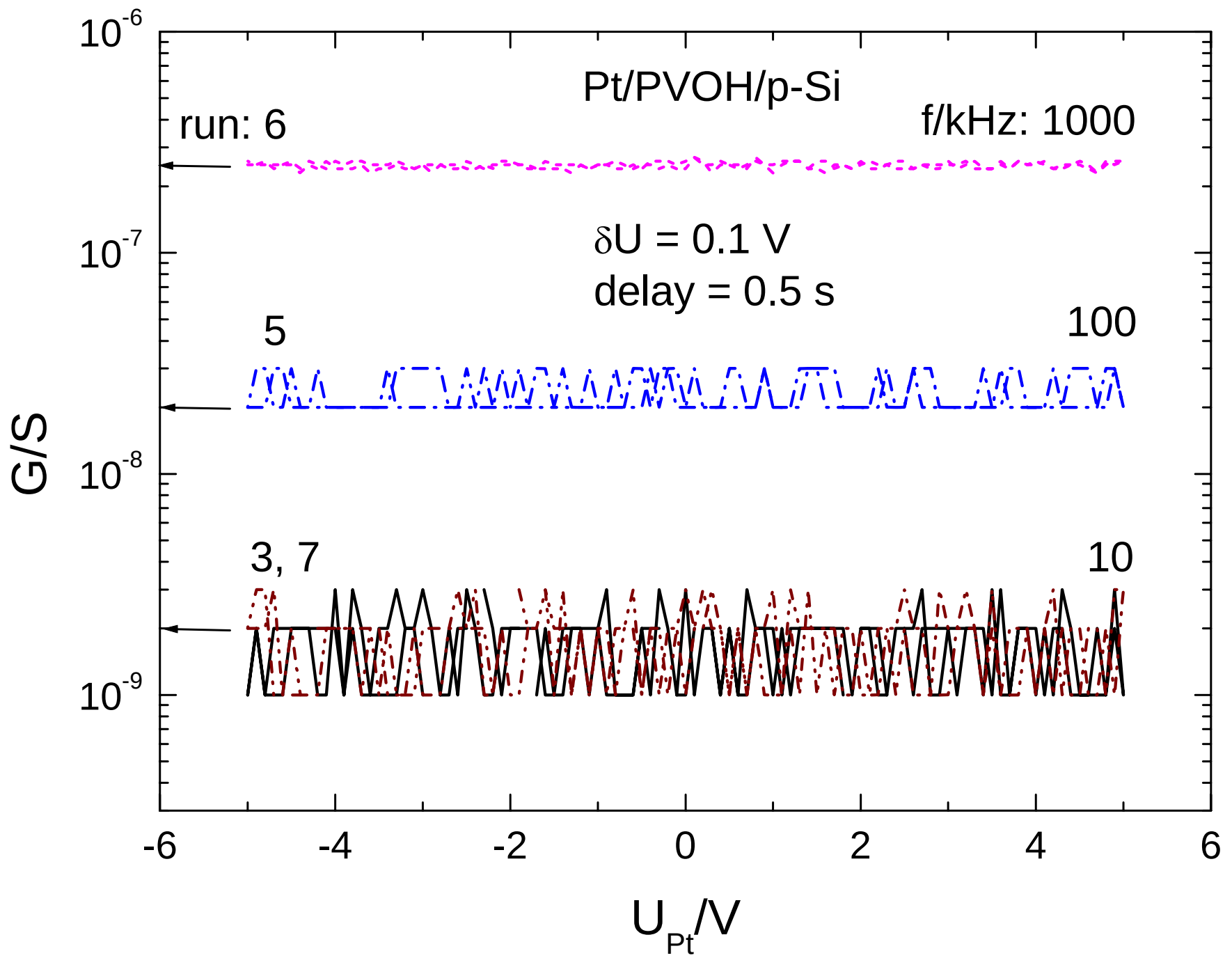

Fig. 7

**References**

[1] G.H. Gelinick, T.C.TGeunus, D.M. de Leeuw: Appl.Phys.Lett. 77(10)(2000), p.1487
[2] H.Sirringhaus, N.Tessler, R.H. Friend: Synthetic Metals, 103(1999), p.857.
[3] K. Maexa,M. R. Baklanov, D. Shamiryan and F. Iacopi, S. H. Brongersma,Z. S. Yanovitskaya, Low dielectric constant materials for microelectronics: Journal of Applied Physics Vol. 93, Number 11,1 June 2003.
[4] Janos Veres, Simon D.Ogier, Stephen W. Leeming, Domenico C. Cupertino and Soad Mohialdin Khaffaf: Adv.Funct. Mater. 2003, 13, No 3, March.
[5] Müller K, Bär W ,Henkel K, Jahnke A, Schwiertz C, Schmeißer D : Oldenburg Wissenschaftsverlag, 70, 12/2003, p. 565 -568.
[6] http://www.clo.cam.ac.uk/horizon/horizondocs/RichardFriend.pdf
[7] Bao Z et al. Chem Mater 1997;9: 619.
[8] Knobloch A, Bernds A, Clemens W. IEEE session 4: Polymer Electronics Devices.
[9] Garnier F, Hajlaoui R, Yassar A, Srivastava P: Science 1994, 265.
[10] Tate J, Rogers JA, Jones CDW:Langmir 2000, 16, 6054.
[11] Bao Z, Rogers JA, Katz Howard EJ.: Mater Chem, 1999, 9, 1895- 904.
[12] Chen B, Cui T, Liu Y, Varahranyan K: Solid- State Electronics, 47, 2003, 841-847.
[13] Kawas T, Sirrunhaus H,Friend RH, Shimoda T: IEEE Internationl Electronics Meeting, December 10-13, 2000.
[14] Sirrunhaus H, Kawas T, H,Friend RH, Shimoda T, Inbasekaran M, Wu W:Science, 2000, 290, 2123.
[15] Knobloch A, Bernfs A, Clemens W, Printed Polymer Transistors.
[16] Liu Y, Cui T, Varahranyan K: Solid- State Electronics, 47, 2003, 811-814.
[17] HP Model 4061A Semiconductor/Component Test System.
[18] Johannes Dechant, Ultrarotspektroskopische Untersuchungen an Polymeren, Akademie-Verlag, Berlin, 1972
[19] Brütting W, Riel H, Beierlein T, Riess W, J. Appl. Phys. (2001), 89,1704.
[20] Thurzo I, Pham G, Zahn D R T, Chem. Phys. 2003, 287, 43.
[21] León C, Lucía M L., Santamaría J, Phys. Rev. B 1997, 55, 882.
[22] Ngai K L, León C, Phys. Rev. B 1999, 60, 9396.
[23] Macdonald J R, J. Chem. Phys. 2001,115, 6192.
[24] León C, Rivera A, Várez A, Sanz J, Santamaría J, Phys. Rev. Lett. 2001 86, 1279.
[25] Rivera A, León C, Sanz J, Santamaría J, Moynihan C T, Ngai K L, Phys. Rev. B 2002, 65, 224302.
[26] León C, Rivera A, Várez A, Sanz J, Santamaría J, Moynihan C T, Ngai K L, J. Non-Cryst. Sol. 2002, 305, 88
[27] Macdonald J R, J. Chem. Phys. (2002), 116, 3401.

.